\documentstyle[12pt,epsf]{article}

\begin{document}
\def \cosw{\cos \chi_w^t}
\def \sinw{\sin \chi_w^t}
\def \ce{\cos \chi_e^w}
\def \se{\sin \chi_e^w}
\def \sse{\sin^2 \chi_e^w}
\def \pew{\Phi}
\def \chie{\chi_e^w}
\def \chiw{\chi_w^t}
\def \sw{\sin \theta_W}
\def \ssqw{\sin^2 \theta_W}
\def \cw{\cos \theta_W}
\def \thetad{\theta^{\dagger}}
\def \thetas{\theta^*}
\def \sangle{\xi}
\def \ds{\frac{d \sigma}{d \cos \theta^*}}
\def \norm{ \Bigl( ~\frac{3 \pi \alpha^2}{2 S}\beta ~\Bigr)~}
\def \norme{\Bigl( ~\frac{3 \pi \alpha^2}{8 s}\beta ~\Bigr)~}
\def \norms{\Bigl( ~\frac{ \pi \alpha_s^2}{9 s}\beta ~\Bigr)~}
\def \beq{\begin{equation}}
\def \eeq{\end{equation}}
\def \beqa{\begin{eqnarray}}
\def \eeqa{\end{eqnarray}}
\def \amp{{\cal M}}
\def \ttbar{$t \bar t$}
\def \qqbar{$q \bar q$}
\def \antibeamline{``anti-beamline''}
\def \beamline{Beamline}
\def \UP{{\rm R}}
\def \DOWN{{\rm L}}
\def \up{\uparrow}
\def \down{\downarrow}
\def \half{\hbox{$1\over2$}}
\def \qbar{\bar q}
\def \tbar{\bar t}
\def \ebar{\bar e}
\def \opt{Off-diagonal}
\def \epm{$e^+e^-$}
\def \eL{$e^-_L~e^+_R$}
\def \eR{$e^-_R~e^+_L$}
\def \UD{$t_{\up} ~{\bar t}_{\down}$ }
\def \DU{$t_{\down} ~{\bar t}_{\up}$ }
\def \UU{$t_{\up} ~{\bar t}_{\up}$ }
\def \DD{$t_{\down} ~{\bar t}_{\down}$ }


\def	\nn		{\nonumber}
\def	\=		{\;=\;} 
\def	\ret		{\\[\eqskip]}
\def	\to		{\rightarrow }
 
\begin{flushright}
\vbox{\begin{tabular}{l} 
        FERMILAB-Conf-98/212-T \\
	hep-ph/9807573\\
        July 31, 1998
        \end{tabular}}
\end{flushright}
 
\begin{center}
\vspace{+1cm}
\Large
\Large{Why ${\bf e^+e^- \rightarrow t\bar{t}}$ is 
different}\footnote{Dedicated to Sid Drell on the occasion
of the ``Sid Drell Symposium''
held at the Stanford Linear Accelerator Center, Stanford University,
California on July 31, 1998.}
\\ \vskip 0.5cm
\large
         Stephen Parke \\
\vskip 0.1cm
{\small \em Theoretical Physics Department \\
Fermi National Accelerator Laboratory \\ 
P.O. Box 500, Batavia, IL 60510, U.S.A.\\
E-mail: parke@fnal.gov}
\end{center}
\thispagestyle{empty}
\vskip 0.7cm
 
\begin{abstract}
Unlike other examples of fermion pair production in \epm\ collisions,
we show that top-quark pairs are 
produced in an essentially unique spin configuration
in polarized $e^+e^-$ colliders at all energies.
Since the directions of the electroweak  decay products 
of polarized top-quarks are strongly
correlated to the top-quark spin axis, this unique spin configuration
leads to a  distinctive topology for top-quark pair events 
which can be used to 
constrain anomalous couplings to the top-quark.
A significant interference effect between the {\it longitudinal} and 
{\it transverse} W-bosons in the decay of polarized top-quarks
is also discussed.
\end{abstract}
\newpage

At an $e^+e^-$ collider with center of mass energy from
400 to 1000 GeV, top quark~\cite{CDFD0} pair production will
be very different than fermion pair production at any previous $e^+e^-$
machine for the following reasons:
the photon ($\gamma$), $Z$-boson and their interference
will contribute 
approximately equally to the production and the top quark pairs 
will be produced at non-ultra-relativistic speeds. 
At present and previous $e^+e^-$ colliders fermion pair production is 
either in the ultra-relativistic regime (LEP, SLC and TRISTAN) 
and/or completely dominated by the photon contribution 
(PETRA, PEP, CESR, SPEAR and CEA).
This marked difference in the production mechanism leads to significantly
different correlations, both angular and in spin, 
of the top and anti-top quarks compared with any previous fermion
pairs produced in $e^+e^-$ collisions.
Moreover, even the spin correlations are measurable for
top quark pair production because the top quark decays~\cite{Bigi} 
before QCD effects
de-correlate its spin and the decay products of a top
quark are highly correlated with the direction of its spin
{\it i.e. the top quark behaves more like a charged lepton than
one of the light quarks in this regard.}

The matrix element for $e^-_Le^+_R \rightarrow t_s \bar{t}_{\bar{s}}$ is
\beq
{\cal{M}} \sim {e^2 \over S} 
~\bar{\bf v}(\bar{e})\gamma^{\mu} \gamma_L {\bf u}(e) 
~~\bar{\bf u}(t,s) \gamma_{\mu}
\{~f_{LL}  \gamma_L ~+~f_{LR} \gamma_R ~\}
{\bf v}(\bar{t},\bar{s}) 
\label{MatrixElement}
\eeq
where $\gamma_{R,L} = \frac{(1\pm \gamma_5)}{2}$, $\sqrt{S}$
is the center of mass energy and $s$ and $\bar{s}$ are the spins
of the top and anti-top quarks. 
The $f_{IJ}$'s 
are the sum of the photon and $Z$-boson products of couplings
to the fermions
corrected for the difference in the propogators.
For top production the $f_{IJ}$'s have a very weak S dependence 
and are given by
\beq
f_{LL}= 1.22 - 1.19 \quad f_{LR}= 0.418 - 0.434
\quad {f_{LR} \over f_{LL}} = 0.343 - 0.365
\eeq
from threshold to ultra-high energies.
Without the $Z$-boson couplings, $f_{LL}=f_{LR}=2/3$ so that the 
contribution from the $Z$-boson is constructive for $f_{LL}$ and
destructive for $f_{LR}$.
This difference between $f_{LL}$ and $f_{LR}$ 
and the fact top quark pairs are non-ultra-relativistic
is what makes top pair production 
different than other examples of fermion pair production.

We will first consider the limit where $f_{LR} = 0$.
In the ultra-high energy limit, where the mass of the top quark,
$m_t$, can be
neglected $\sqrt{S} >>> 2m_t$, the matrix element is  
\beq
{\cal{M}} \sim e^2  ~f_{LL} (1 + \cos \theta^*),
\label{ultraS}
\eeq
 where $\theta^*$ is the scattering angle.
In this limit the top quark is purely left-handed and the anti-top quark
is purely right-handed, $e^-_Le^+_R \rightarrow t_L \bar{t}_R$.

At threshold, $\sqrt{S} \sim 2m_t$, the matrix element is simply
\beq
{\cal{M}} \sim e^2 ~f_{LL} 
\label{thresholdS}
\eeq
and the direction of the top and anti-top quark spins
are opposite the
electron momentum or equivalently in the direction of the positron
momentum.

At intermediate energies the matrix element interpolates between these two
extremes 
\beq
{\cal{M}} \sim e^2 ~f_{LL} (1 + \beta \cos \theta^*),
\label{intermediateS}
\eeq
where $\beta$ is the ZMF speed of the top quarks.
What are the directions of the top and anti-top quark spins?
In the rest frame of the top quark there are three
natural possibilities for the
direction of the top spin;
the electron, the positron or
the anti-top quark momentum directions.
The threshold result excludes the anti-top quark momentum direction,
which is undefined at threshold,
leaving only the electron or positron momentum direction.
Similarly the natural possibilities for 
the anti-top quark spin, in its rest frame, are the electron or
positron momentum directions.
The correct choice is 
{\it that the top quark spin vector is in the direction of the
positron momentum in the top quark rest frame and
the anti-top quark spin vector is in the direction opposite that of the
electron momentum in the anti-top quark rest frame. } 
This spin basis has been called the beamline basis~\cite{MP2}
and smoothly interpolates the required basis 
from threshold to ultra-high energies.
An obvious question is, why is the spin of the top quark associated with
the positron and the anti-top associated with the electron instead of vice
verse.
If the term proportional to $f_{LL}$ in the matrix element,
eqn (\ref{MatrixElement}), is
Fierz re-arranged
one obtains
\beq
\sim ~f_{LL}  ~[~\bar{\bf u}(t,s) \gamma_R {\bf v}(\bar{e})~]  
~[~\bar{\bf u}(e) \gamma_L {\bf v}(\bar{t},\bar{s})~].
\eeq
That is, the top quark spin is associated with the positron and anti-top quark
spin with the electron\footnote{If 
the $f_{LR}$ term of the matrix element, eqn (\ref{MatrixElement}),
was dominant then the top
quark spin would be associated with the electron 
and the anti-top quark with the positron.}.

Returning to non-zero $f_{LR}$, the differential cross-sections
obtained by explicit calculation,
using the positron direction for the top quark spin 
and the electron direction for the
anti-top quark spin in their respective rest frames, are
\beqa
\ds ~(e^-_L~e^+_R \rightarrow t_{\up}~\tbar_{\down} ) & = &
\norm \Bigl[ f_{LL}(1+\beta \cos \thetas)
+f_{LR}\frac{(1-\beta^2)}{(1+\beta\cos\thetas)} \Bigr]^2 , \nn \\[0.1in]
 \ds ~(e^-_L~e^+_R \rightarrow t_{\up}~\tbar_{\up})  & = &
\ds ~(e^-_L~e^+_R \rightarrow t_{\down}~\tbar_{\down}) \nn  \\[0.1in]
& = & \norm
f^2_{LR} ~\frac{\beta^2(1-\beta^2) \sin^2 \thetas }
{(1+\beta \cos \thetas)^2} 
\ ,  \\[0.1in]
\ds ~(e^-_L~e^+_R \rightarrow t_{\down}~\tbar_{\up} ) & = & 
\norm f^2_{LR}~\frac{\beta^4 \sin^4 \thetas}{(1+\beta \cos \thetas)^2} 
\ . \nn
\label{Beamline}
\eeqa
Note only the Up-Down spin configuration is non-zero if $f_{LR}=0$,
confirming
the previous analysis.
The Up-Up and Down-Down components are equal because CP is conserved both
in the physics at this order of perturbation theory and in this spin basis.
The $sin^2 \theta^*$ factor in the Up-Up and Down-Down differential
cross section implies that this is a P-wave and higher contribution.
For the Down-Up component we have a $sin^4 \theta^*$ indicative
of a contribution starting at D-wave.
At ultra-high energies only the Up-Down and Down-Up components are
non-zero, giving the usual helicity basis result~\cite{peskin}.
At threshold only the Up-Down component survives as expected.

It should not be a surprise that by making small changes
to the spin basis for small
non-zero $f_{LR}$,
the contributions from the two terms in the matrix element,
eqn (\ref{MatrixElement}), can be made to 
totally destructively interfere for the Up-Up and Down-Down components.
This spin basis has been called the Off-diagonal basis~\cite{PS}
and for small values of the ratio $f_{LR}/f_{LL}$ gives the following
differential cross sections;

\beqa
\ds ~(e^-_L~e^+_R \rightarrow t_{\up}~\tbar_{\down} ) & = &
\norm  f^2_{LL}(1+\beta \cos \thetas)^2
~(1+{\cal{O}}( \frac{f_{LR}}{f_{LL}}))\ , \nn \\[0.1in]
 \ds ~(e^-_L~e^+_R \rightarrow t_{\up}~\tbar_{\up})  & = &
\ds ~(e^-_L~e^+_R \rightarrow t_{\down}~\tbar_{\down}) ~ \equiv ~0
\ , \\[0.1in]
\label{ApprxOffDiag}
\ds ~(e^-_L~e^+_R \rightarrow t_{\down}~\tbar_{\up} ) & = & 
\norm f^2_{LR}~\frac{\beta^4 \sin^4 \thetas}{(1+\beta \cos \thetas)^2} 
~(1+{\cal{O}}(\frac{f_{LR}}{f_{LL}}))
\ . \nn
\eeqa

\begin{figure}[t]
\epsfysize=3.35in
\centerline{\vbox{\epsfbox{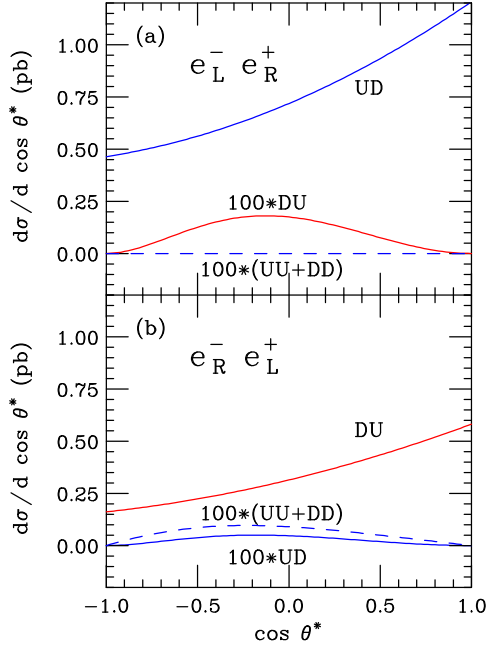}}}
\caption[]{
The spin configurations using the off-diagonal basis for both
$e^-_L~e^+_R$ and $e^-_R~e^+_L$ for
$\sqrt{s}=400~GeV$.
Note that the sub-leading configurations have been amplified 
by a factor of 100 in these figures.
}
\label{offdiag}
\end{figure}

\begin{figure}[b]
\epsfysize=3.35in
\centerline{\vbox{\epsfbox{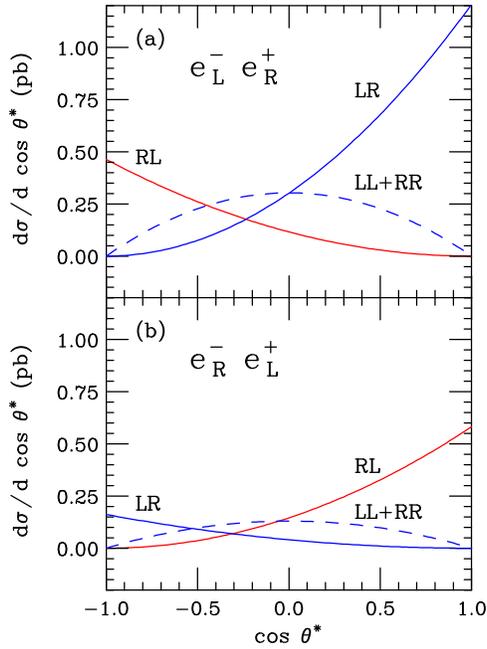}}}
\caption[]{
The spin configurations using the helicity basis for a 
$\sqrt{s}=400~GeV$ collider.
}
\label{hel}
\end{figure}

The Down-Up component is suppress relative to the Up-Down component
by $(f_{LR}/f_{LL})^2$ as well as $\beta^4$ .
The factor $(f_{LR}/f_{LL})^2$ gives an extra order of magnitude suppression
to the Down-Up cross section for top production compared to
examples where $f_{LL}=f_{LR}$.
For a collider of $\sqrt{s}=400$ GeV the top quark pairs have a ZMF speed 
equal to 0.5c and the Up-Down component is 
99.88\% of the total cross section, see Fig.~\ref{offdiag}(a).
In the helicity basis the LR component is only 52\% of the total cross
section at this energy, Fig.~\ref{hel}(a).

For $e^+_Le^-_R$ scattering\footnote{
The couplings are 
$f_{RR}= 0.882 - 0.868 \quad f_{RL}= 0.185 - 0.217
\quad {f_{RL}/f_{RR}} = 0.210 - 0.250$
from threshold to ultra-high energies respectively.}
the same Off-diagonal spin basis can be used
because $f_{RL}/f_{RR} \approx f_{LR}/f_{LL}$ and similar results are
obtained except that the dominant spin component is now Down-Up,
see Fig.~\ref{offdiag}(b) and \ref{hel}(b).

\clearpage

What about the stability of this result under QCD radiative corrections?
The quick answer is that the ${\cal{O}}(\alpha_s)$ corrections are
dominated by the soft gluon emission diagrams which factorize into
an eikonal factor times the tree level result. 
That is, there is no spin flip from soft gluon emission and that only
hard gluon emission or the $\gamma/Z$ anomalous magnetic moment term
can change the tree level result.
An explicit calculation~\cite{KNP} 
of the ${\cal{O}}(\alpha_s)$  corrections gives 
the ratio of the dominant top quark spin to the total as
\beq
{\sigma(e^-_Le^+_R \rightarrow t_{\uparrow} +X)
\over
\sigma(e^-_Le^+_R \rightarrow t +X)}
= 99.85\%
\eeq
at $\sqrt{s}=400$ GeV. This is a small change from the tree level result
of 99.88\%  even though 
the ${\cal{O}}(\alpha_s)$ correction to the total cross section is 28\% !

\begin{figure}[t]
\epsfysize=3.35in
\centerline{\vbox{\epsfbox{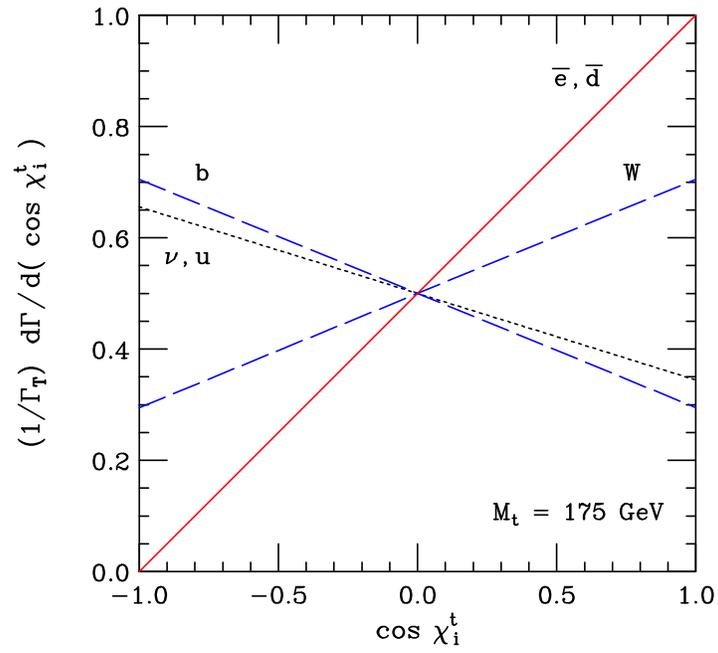}}}
\vspace{0.25cm}
\caption[]{
Correlations of the decay products of the top quark with the spin of the
top quark.
}
\label{one_corr}
\end{figure}

Since the top-quark pairs are produced in an unique spin configuration,
and the electroweak decay products of polarized top-quarks are 
strongly correlated to the spin axis, 
the top-quark events at  \epm\ collider have a
very distinctive topology.
The predominant decay mode of the top-quark is 
$t \rightarrow b W^+$, with the $W^+$ 
decaying either hadronically or leptonically. 
Let us first consider the single particle decay products correlations 
with the top-quark spin.
If $~\chi^t_i~$ is the angle between the top quark spin 
and the momentum of the $i$-th decay product
measured in the top-quark rest-frame
then the differential decay rate of the top-quark is
\beqa
\frac{1}{\Gamma_T} ~{d ~\Gamma \over d \cos \chi^t_i} & = &
\frac{1}{2}
~\Bigl[1 + \alpha_i \cos \chi^t_i \Bigr] \ ,
\eeqa
where 
\beqa 
\alpha_{e^+} & = & \alpha_{\bar{d}}  \equiv 1 \nn \\
\alpha_W & =  & -\alpha_b  = \frac{m^2_t - 2 m^2_W}{m^2_t+2 m^2_W} 
\approx 0.41 \nn \\
\alpha_{\nu} & = &  \alpha_u  \approx -0.31 \nn
\eeqa
see Je\.zabek and K\"uhn~\cite{Jezabek2}.
Fig.~\ref{one_corr} shows these single particle correlations.
\begin{figure}[b]
\epsfysize=3.35in
\centerline{\vbox{\epsfbox{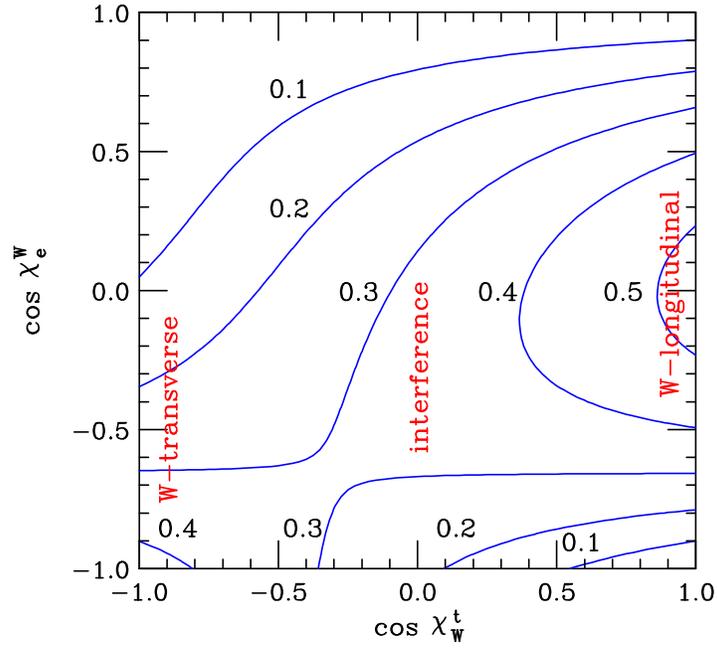}}}
\vspace{0.25cm}
\caption[]{
Contours of the top-quark differential angular decay distribution,
$\frac{1}{\Gamma_T} ~{d^2 ~\Gamma \over d \ce ~d \cosw }$,
in the  $\cos \chie-\cos \chiw$ plane. 
}
\label{two_corr}
\end{figure}
\clearpage
Note that the positron or d-quark are 
maximally correlated with the top quark spin.
What is surprising about this is that the charged lepton or d-quark
came from the decay of the $W$-boson 
which is less correlated with the spin of the top quark.
This maximal correlation of the charged lepton or d-quark 
requires a subtle cancelation between the amplitudes involving 
transverse and longitudinal $W$-boson in 
Standard Model top quark decay\cite{PS}.

Second, there are significant two particle correlations.
Fig.~\ref{two_corr} shows contour plots of the differential angular decay distribution
in the $\cos \chie$ verses $\cos \chiw$ plane
where $\pi - \chie$ is the angle between the direction of 
motion of the $b$-quark and the positron (or d-quark) in the W-boson rest-frame. 
When the $W$-boson momentum is parallel to the top spin, 
$\cos \chiw = 1$, the $W$-boson is purely longitudinal as can be seen 
from the $\cos \chie$ distribution of this figure along the right hand edge. 
Whereas, when the $W$-boson momentum is anti-parallel to the top spin,
$\cos \chiw=-1$, the $W$-bosons is purely transverse (left-handed).
In between these two extremes both longitudinal and transverse $W$-boson
contribute and there is significant interference effects between these
two amplitudes.

In this paper we have shown that top quark pair production at an
\epm\ collider is very different than fermion pair production at current and past
\epm\ machines. 
At a polarized \epm\ collider the top quark pairs are produced in an essentially 
unique spin configuration.
In this configuration, the top-quark spin is strongly correlated with 
the positron spin direction determined in the top-quark rest-frame.
The subsequent electroweak decays of the top-quark pair give decay 
products whose angular distributions are highly correlated with the parent
top-quark spin. 
Top-quark pair events thus have a distinctive topology.
This topology is sensitive to  the  top-quark couplings to the $Z$-boson
and to the photon, which determine    
the orientation and the size of the top-quark and top anti-quark 
polarizations,  
as well as to the top-quark couplings to the $W$ and the $b$-quark, 
which determine its decay distributions. 
Angular correlations in top-quark events 
may therefore be used to constrain deviations 
from the Standard Model.
We have also shown that the interference between the {\it longitudinal}
and {\it transverse}
$W$-bosons has a significant impact on the angular distribution 
of the top-quark decay products, and thus will 
provide additional means for 
testing the Standard Model predictions for top-quark decays.


\section*{Acknowledgments}
The author would like to take this opportunity to thank Sid Drell
for many stimulating physics discussions over the years.
The Fermi National Accelerator 
Laboratory is operated by Universities Research Association,
Inc., under contract DE-AC02-76CHO3000 with the U.S. Department
of Energy.


\end{document}